\documentclass[conference]{IEEEtran}
\IEEEoverridecommandlockouts
\usepackage{cite}
\usepackage{amsmath,amssymb,amsfonts}
\usepackage{algorithmic}
\usepackage{hyperref}
\usepackage{cleveref}
\usepackage{graphicx}
\usepackage{caption}
\usepackage{subcaption}
\usepackage{amsmath}
\usepackage{adjustbox}
\usepackage{tcolorbox}
\usepackage{textcomp}
\usepackage{xcolor}
\usepackage{hyperref}
\def\BibTeX{{\rm B\kern-.05em{\sc i\kern-.025em b}\kern-.08em
    T\kern-.1667em\lower.7ex\hbox{E}\kern-.125emX}}
    
\begin{document}

\title{A Framework for Energy-aware Evaluation of Distributed Data Processing Platforms in Edge-Cloud Environment\\
% {\footnotesize \textsuperscript{*}Note: Sub-titles are not captured in Xplore and
% should not be used}
}

\author{\IEEEauthorblockN{Faheem Ullah, Imaduddin Mohammed, Ali Babar}
\IEEEauthorblockA{
% \textit{CREST\textsuperscript{1}\thanks{\noindent\textsuperscript{1}Centre for Research on Engineering Software Technologies}, The University of Adelaide}\\
School of Computer Science, The University of Adelaide, Australia\\
\{faheem.ullah, imaduddin.mohammed, ali.babar\}@adelaide.edu.au}
% \and
% \IEEEauthorblockN{Shagun Dhingra}
% \IEEEauthorblockA{\textit{School of Computer Science} \\
% \textit{CREST, The University of Adelaide}\\
% Adelaide, Australia\\
% shagun.dhingra@adelaide.edu.au}
% \and
% \IEEEauthorblockN{M. Ali Babar}
% \IEEEauthorblockA{\textit{School of Computer Science} \\
% \textit{CREST, The University of Adelaide}\\
% Adelaide, Australia\\
% ali.babar@adelaide.edu.au}
}
\maketitle
\thispagestyle{plain}
\pagestyle{plain}
\begin{abstract}
Distributed data processing platforms (e.g., Hadoop, Spark, and Flink) are widely used to distribute the storage and processing of data among computing nodes of a cloud. The centralization of cloud resources has given birth to edge computing, which enables the processing of data closer to the data source instead of sending to the cloud. However, due to resource constraints such as energy limitations, edge computing cannot be used for deploying all kinds of applications. Therefore, tasks are offloaded from an edge device to the more resourceful cloud. Previous research has evaluated the energy consumption of the distributed data processing platforms in isolated cloud and edge environments. However, there is a paucity of research on evaluating the energy consumption of these platforms in an integrated edge-cloud environment, where tasks are offloaded from a resource-constraint device to a resource-rich device. Therefore, in this paper, we first present a framework for the energy-aware evaluation of the distributed data processing platforms. We then leverage the proposed framework to evaluate the energy consumption of the three most widely used platforms (i.e., Hadoop, Spark, and Flink) in an integrated edge-cloud environment consisting of Raspberry Pi, edge node, edge server node, private cloud, and public cloud. Our evaluation reveals that (i) Flink is most energy-efficient followed by Spark and Hadoop is found least energy-efficient (ii) offloading tasks from resource-constraint to resource-rich devices reduces energy consumption by 55.2\%, and (iii) bandwidth and distance between client and server are found key factors impacting the energy consumption.  
\end{abstract}

\begin{IEEEkeywords}
Edge computing, Cloud computing, Energy Consumption, Task Offloading, Hadoop, Spark, Flink.
\end{IEEEkeywords}

\section{Introduction}\label{intro}
The centralization of computing nodes in a cloud introduces significant delay especially for data-intensive tasks \cite{wang2020survey}. This is because data needs to be transmitted from data source to a centralized cloud via Wide Area Network (WAN) with a limited bandwidth and high latency. Moreover, transmitting data from a data source located at a distance from a cloud may lead to network congestion. The downsides of the centralized cloud have given birth to a new paradigm, \textit{edge computing}, which stores and processes data closer to a data source instead of sending it to a resource-rich cloud \cite{xu2019computation}. As a result, the response time improves and the bandwidth consumption reduces. However, edge computing faces its own set of challenges. First, edge devices such as sensors have limited storage and computation capability. Second, with heavy load, the energy consumption of an edge device increases, which leads to quicker depletion of the limited battery life of edge device.  To cater the shortcomings of pure cloud computing and edge computing, the concept of \textit{task offloading} is coined, where a task is offloaded from the resource-constrained edge device to the resource-rich cloud in an integrated edge-cloud environment \cite{jiang2019toward}. However, not all tasks can always be offloaded as it will nullify the main concept of edge computing that warrants to execute tasks at the edge. Hence, a question arises \textit{"Where, what, how, and when to offload?"}. 

In order to answer \textit{'Where to offload'} (the focus of this paper), there are several options such as edge node, adjacent server, or cloud. The answer to this question primarily depends upon the type of task and the network layout of the edge-cloud environment. The key types of tasks include data processing and data storage. Data processing in a distributed environment is carried out using distributed data processing platforms (hereafter referred as data platforms), which distribute the storage and processing of data among computing nodes. The most commonly used platforms include Apache Hadoop, Apache Spark, and Apache Flink \cite{30}. These platforms differ from each other in several ways. For example, Hadoop is a disk-based framework while Spark and Flink are memory-based frameworks. Also, Hadoop is designed for batch processing whereas Spark and Flink are designed both for batch and stream processing. Such differences with respect to the architecture and features directly impact their response time, scalability, and energy consumption. 

A plethora of research exists on the evaluation of these platforms in cloud as well as edge environment \cite{xu2019computation},\cite{30},\cite{komninos2019performance}. However, none of the existing studies have evaluated these platforms in an integrated edge-cloud environment, where data processing tasks are offloaded from a resource-constrained device to a resource-rich device. Moreover, these platforms are often evaluated in terms of response time and scalability \cite{singh2015survey},\cite{ali2019survey}. While response time and scalability are important metrics in cloud computing, they alone are not enough in the context of edge-cloud environment. This is because in addition to response time and scalability, energy consumption also becomes a key metric in edge-cloud enviornment. The reason for this is that edge devices, such as mobile sensors, have limited battery life. Applications running on edge devices that consume too much energy deplete the battery quickly. Consequently, a user has to recharge the edge device on frequent basis, which results in poor user experience. Hence, it is important to answer the question - \textbf{\textit{How energy-efficient are various distributed data processing platforms in an edge-cloud environment?}}

In this paper, we answer this question by first presenting a framework for the energy-aware evaluation of the data platforms in an edge-cloud enviornment. We then use the proposed framework to evaluate the energy consumption of the three widely used platforms (i.e., Hadoop, Spark, and Flink) during task offloading from resource-constraint to resource-rich devices. According to our evaluation framework, we have designed different usage scenarios based on client and server nodes on which these data platforms are deployed. In our setup, client node offloads a task and server node executes the offloaded task.  Our experimental infrastructure consists of five types of computing nodes - Raspberry Pi (RPI), laptop (hereafter termed as edge node), high-performance server (hereafter referred as edge sever node), private cloud (OpenStack), and public cloud (Google). We have considered nine offloading scenarios i.e., RPI $\rightarrow$ (edge node, edge server node, private cloud, public cloud), edge node $\rightarrow$ (edge server node, private cloud, public cloud), and edge server node $\rightarrow$ (private cloud, public cloud). The node on the left of the arrow specifies the client node that offloads tasks while the node on the right of the arrow specifies the server node that executes the offloaded task\footnote{Throughout the text and figures/tables, \textit{(x, y)} specifies the scenario where \textit{x} and \textit{y} denote client node and server node, respectively.}. In addition to these scenarios, for the sake of comparison, we have also considered baseline/non-offloading scenarios, where the node offloading a task is the same as the node executing the task i.e., the task is executed locally. We measure and compare the energy consumption of these computing nodes during various stages such as data generation, data transmission, and data processing. In a nutshell, we make the following contributions in this paper.

\begin{itemize}

    \item We propose a framework for the energy-aware evaluation of the distributed data processing platforms in an integrated edge-cloud enviornment.
    \item We demonstrate the use of the proposed framework to evaluate the energy consumption of the three most widely used distributed data processing platforms (Hadoop, Spark, and Flink) during several task-offloading and baseline/non-offloading scenarios.
\end{itemize}

The rest of the paper is structured as follows. Section \ref{frameworks} introduces the three platforms and delineates the related work. Section \ref{EvaluationFramework}
presents our evaluation framework, which is followed by Section \ref{Results} that presents the evaluation results. Section \ref{conclusion} concludes the paper. 

\section{Background and Related Work}\label{frameworks}
In this section, we first briefly introduce the data platforms and present the related work.

\subsection{Distributed Data Processing Platforms}
Distributed data processing platforms are employed in a 
system for distributing the storage and processing of data 
across a cluster of nodes. In this study, we have selected 
the following three data platforms based on their widespread usage in industry and academia \cite{30},\cite{4},\cite{19}, .

\textbf{Hadoop} is one of the earliest and widely adopted data platform. This is a batch processing platform, which follows the functional 
programming model of MapReduce \cite{1}. Hadoop offers an open-source 
Java implementation of MapReduce. It is composed of two 
layers - a data storage layer called Hadoop Distributed File 
System (HDFS) and a data processing layer - Hadoop 
MapReduce Framework \cite{1}. \textbf{Spark} is a novel data platform, which is designed to overcome the shortcomings of Hadoop (e.g., 
lack of user interface and programming language barriers) \cite{15},\cite{20}. Spark does not have its own distributed file system; therefore, it uses 
HDFS as its input source and output destination. Unlike Hadoop, Spark is a 
memory-based platform. Before an operation, the user 
driver program launches multiple workers. Then, these 
workers read data from a distributed file system and cache 
them in their memory as partitions of Resilient Distributed 
Dataset (RDD). This feature enables Spark to avoid 
reloading data at every iteration, which helps in boosting 
the data processing speed. Also, Spark uses RDDs to rebuild information (lost due to potential failures) from incestor RDDs. \textbf{Flink} is relatively a new open-source platform suitable for both 
batch and stream processing \cite{3}. Flink uses a high 
throughput streaming engine that is written in Java and 
Scala. Similar to Hadoop and Spark, Flink follows a 
master-slave architecture, where a master node manages 
the distribution of data processing among worker nodes 
and worker nodes perform the actual data processing. 
Flink does not provide its own storage; therefore, we have 
to use it in combination with another platform such as 
HDFS. Like Spark, Flink is 
a memory-based framework. However, unlike Spark that implements iterations as \textit{for} loops, Flink executes iterations as cyclic data flows. The iterative process in Flink can significanly speedup certain algorithms as Flink reduces the required work in each subsequent iteration.

\subsection{Related Work}
We categorize the related work into the following three sections to position the novelty of our work with respect to the related works.

\subsubsection{Energy consumption of data platforms in cloud}
Several studies explored the energy consumption of data platforms in a cloud environment. Leverich et al. \cite{leverich2010energy} proposed an approach for optimizing the energy consumption of Hadoop. The proposed approach shifts data processing tasks from low-power nodes to high-power nodes. Hence, some nodes can be shutdown to reduce energy consumption. The authors show that such approach saves 9 - 50\% energy consumption. Feller et al. \cite{feller2015performance} studied the energy consumption of two Hadoop clusters. In the first cluster, data resides on the same node that is supposed to process the data while in the second cluster data resides on a different node than the one that is supposed to process the data. It was found that energy consumption significantly decreases if data is placed on the same node that processes the data. Kaushik and Bhandarkar \cite{kaushik2010greenhdfs} proposed GreenHDFS - a variant of Hadoop that places the data in hot and cold zones in a way that the data in cold zone is rarely accessed. Hence, the servers hosting the cold zone largely remain idle, which results in saving 26\% energy. Maroulis et al. \cite{maroulis2017framework} presented a tuning approach for Spark to reduce energy consumption with minimum impact on the execution time. The approach tunes the allocation of the number of CPU cores and CPU frequency to save energy. Experimental evaluation revealed 61 - 75\% energy saving. In \cite{maroulis2019holistic}, the authors proposed ExpREsS - a scheduler that executes Spark applications in a way to minimize energy consumption. This scheduler orchestrates the execution workflow of Spark jobs based on energy predictions obtained from regression analysis. Chatterjee and Morin \cite{chatterjee2018experimental} contrasted the energy consumption of Flink with Storm and Heron in a 8-node cluster. It was observed that Flink outclasses both Storm and Heron in terms of energy consumption in a cloud environment. Veiga et al. \cite{veiga2018bdev} compared the energy consumption of Hadoop, Spark, and Flink deployed on a 16-node cluster. It was found that Hadoop is more energy-efficient for batch workloads while Spark and Flink were found to be more energy-efficient for iterative workloads. All these existing works study the energy consumption of data platforms in a pure cloud environment with no task-offloading. Differently, our work explores the energy consumption of data platforms during task-offloading in an integrated edge-cloud environment. 

\subsubsection{Data platforms in edge computing}
Several studies have explored the use of data platforms in edge computing. For instance, to reduce transmission energy, Dong al. \cite{dong2019computation} proposed a graph-based approach for computation offloading that determines which tasks to process locally on an edge device and which tasks to offload to a cloud server. Spark was installed on both edge and cloud devices. Experimental results show significant reduction in energy consumption during computation offloading. Unlike \cite{dong2019computation} which focuses on what task to offload, our work explores about where to offload. Focusing on the use of Spark for edge computing, Hajji and Tso et al. \cite{hajji2016understanding} evaluated the execution time, CPU usage, and energy consumption of Spark deployed on a single RPI and a cluster of 12 RPI's. It was observed that the energy consumption largely depends upon the workload i.e., more intensive the workload, higher is the energy consumption. In \cite{hajji2016understanding}, the focus is on a standalone RPI or RPI cluster unlike our work where we focus on task offloading. Scolati et al. \cite{scolati2019containerized} contrasted the performance of Hadoop and Spark in edge environment, however, on a containerized cluster of RPI's. In this study, the authors measured CPU and RAM usage to report the limitations of RPI for the deployment of Hadoop and Spark. Komninos et al. \cite{komninos2019performance} investigated the use of RPI cluster in an edge computing scenario for performing machine learning tasks. The authors used Spark for running a machine learning job on a distributed set of RPI nodes (Raspberry Pi Model 4). The results show that the cluster took 50 - 650 seconds to train the model and 0.1 - 280 seconds to test the model. Our work differs from the above works in terms of evaluation metrics, infrastructure model, and data platforms. 

\subsubsection{Energy consumption in task offloading} A plethora of research exists on exploring energy consumption during task offloading \cite{wang2020survey}. Xu et al. \cite{xu2019computation} proposed a multi-objective optimization method with an aim to optimize execution time and energy consumption of the mobile devices during offloading tasks from a mobile device to a cloud. Mansouri et al. \cite{mansouri2021evaluation} studied the energy consumption of distributed databases (i.e., Cassandra, MongoDB, Redis, and MySQL) in hybrid clouds and edge environment. For this purpose, the authors deployed the databases on hybrid cloud (consisting of OpenStack and Azure) and various edge nodes such as RPI and RPI cluster. In \cite{barbera2013offload}, the authors studied whether it is feasible in terms of energy consumption and bandwidth to offload tasks from smartphones to Amazon EC2 public cloud. The Android smartphones were running various Android applications and connected to the cloud via WiFi, 2G, or 3G network. In \cite{li2020energy} and \cite{guo2019toward}, the authors proposed techniques for energy-aware task offloading in edge-cloud environment. The technique proposed in \cite{li2020energy} uses estimation functions to estimate energy and execution time while the one in \cite{guo2019toward} uses machine learning for predicting energy consumption. Based on the estimation, it is decided whether or not to offload tasks. Unlike the previous works, our work focuses on the energy consumption of data platforms in various scenarios to determine where to offload. Hence, our work differs from the previous works in terms of objective, applications, and infrastructure model. 

\section{The Proposed Evaluation Framework}\label{EvaluationFramework}
In this section, we propose our framework for the energy-aware evaluation of the data platforms. The framework consists of seven dimensions, which are reported in the following sub-sections. In this section, we also report on how we leveraged the proposed framework for the energy-aware evaluation of Hadoop, Spark, and Flink in an edge-cloud environment.

\subsection{Evaluation Goal}
Evaluation goal underlines the key reason or motivation behind undertaking an evaluation study. The evaluation goal is defined by the researchers or practitioners. Examples of goals include comparison of the energy consumption of different platforms, selection of the appropriate hardware devices keeping in view the energy constraints, and assessing the energy consumption of new software/hardware solutions. For this study, our goal is to compare the energy consumption of the three data platforms in various task offloading and non-offloading scenarios. 

\subsection{System Architecture}\label{system_architecture}
System architecture visualizes the different elements of the entire infrastructure used for the evaluation. These elements include edge devices, different types of clouds, measurements probes, and software applications. The architecture used in our case study is presented in Fig. \ref{architecture}, which is divided into four components. The bottom component is the \textit{hardware} layer that hosts the following five studied devices - RPI, edge node (laptop), resource-rich server (edge server node), private cloud (OpenStack), and public cloud (Google). These devices are connected with each other via wireless connections. The \textit{application} component underpins the type of job/task to be executed using the platforms deployed on the hardware devices. Some of the popular tasks include machine learning, searching, and sorting. For the sake of rigorous evaluation, we considered several types of tasks in our study, which are described in Section \ref{workloads}. The \textit{data platforms} part specifies the platforms considered for evaluation. These platforms are deployed on the devices in the hardware layer. Whilst several platforms are available in the market, we considered the evaluation of Hadoop, Spark, and Flink in this study based on their widespread use and support from the well-known cloud providers such as Google, AWS, and Azure \cite{singh2015survey}. The \textit{measurements} component specifies the key evaluation metrics which include but are not limited to energy consumption, time consumed in the execution of application workload, and resources (e.g., CPU and RAM) utilized during job execution. Our study primarily focuses on energy consumption, however, also measures other metrics as discussed in Section \ref{evaluation-metrics}.

\begin{figure}[t]
\captionsetup{justification=centering}
\centering
  \begin{subfigure}{0.35\textwidth}
  \centering
  \includegraphics[width=\linewidth]{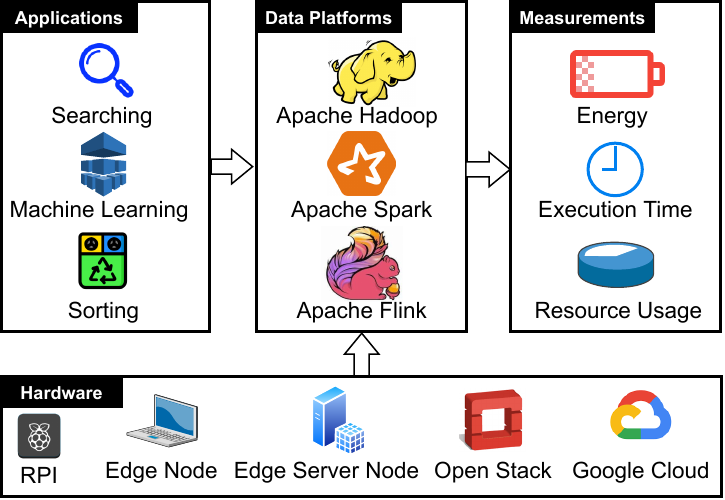}
%   \caption{Flink}
  %\vspace{5.00mm}
  %\label{bandwidth}
\end{subfigure}
\caption{Edge-cloud architecture used in this study.}\label{architecture}
\label{architecture}
\vspace{-1.5 em}
\end{figure}

\subsection{Environmental Parameters}
\begin{figure}[t]
\captionsetup{justification=centering}
\centering
  \begin{subfigure}{0.485\textwidth}
  \centering
  \includegraphics[width=\linewidth]{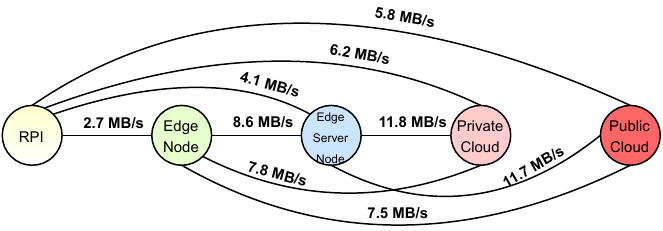}
%   \caption{Flink}
  %\vspace{5.00mm}
  %\label{bandwidth}
\end{subfigure}
\caption{Mean bandwidth (in Megabytes/sec) between the five nodes in our infrastructure.} 
\label{bandwidth}
\vspace{-1.5 em}
\end{figure}

Several parameters in experimental test-bed directly impact the energy consumption. Therefore, these parameters need to be considered during the interpretation of the results. Examples of such parameters include but are not limited to network speed, storage type (e.g., HDD or SSD), and VM flavours. We paid special attention to the network speed as our study involves transferring data from one device to another using network connection. Therefore, we used a networking tool, IPerf3\footnote{https://iperf.fr/}, to measure the bandwidth between different devices considered in our study. We installed Iperf3 on each device and ran it for 10 minutes to measure the bandwidth shown in Fig. \ref{bandwidth}. The bandwidth shown in Fig. \ref{bandwidth} is the average of the bandwidth measured in each second during 10 min. The highest bandwidth is observed between edge server node and private cloud while the lowest is observed between RPI and edge node. Given that we did not consider task offloading from private to public cloud (see Section \ref{scenarios} for details), we did not measure the bandwidth between private and public cloud.

\subsection{Evaluation Workloads}\label{workloads}
As described in Section \ref{system_architecture}, the application layer in Fig. \ref{architecture} specifies various tasks/workloads to be executed using the platforms. These workloads directly impact the energy consumption, therefore, the selected workloads should be representative of the real-world scenarios \cite{veiga2018bdev}. We selected two types of workloads – batch and iterative. Batch workload implements one-pass processing, where the input is processed exactly once. Iterative workload implements loop-caching, where the same input is processed multiple times, however, not necessarily in a streaming fashion. That is why, all four workloads (two of each type) used in this paper are applicable for Hadoop, Spark, and Flink. 

In the batch category, we chose word count and grep, which are common in end-user applications such as google search \cite{17,18}. In the iterative category, we selected K-means and PageRank, which are also frequently used in applications such as machine learning and graph processing \cite{17},\cite{18}. These workloads are also most popular among researchers for evaluation purposes \cite{14},\cite{25}. We generated 3 GB data for each workload. \textit{Grep} is a workload that implements a command for searching plain text generated using Wikipedia text files \cite{15}. \textit{WordCount} counts the number of times a word appears in a text file. It can reasonably assess the aggregation capability of distributed data processing platforms.\cite{15}. \textit{K-means} is the well-known K-means algorithm for clustering data, which is a good choice for assessing the ML feature of a data platform \cite{18}. This workload receives input from a random data generator. \textit{PageRank} is a graph algorithm, which ranks a set of elements according to their references \cite{15}. This workload is representative of search indexing. As input, this workload uses Wikipedia page-to-page link database.

\subsection{Evaluation Metrics}\label{evaluation-metrics}
In the context of energy-aware evaluation, the most important metric is the consumed energy measured in the unit of Joules. However, to understand underlying reasons behind energy consumption, it is important to also measure the execution time and utilization of resources such as CPU, RAM, and disk. The measurement of energy consumption depends upon the hardware and software tools used. For our study, we measured the energy consumption of RPI using USB Power Meter (UPM). UPM provides the required energy readings both in the form of displaying it on the built-in screen as well writing them to a file. For measuring the energy consumption of edge node and edge server node, we leveraged Running Average Power Limit (RAPL), which a a powerful tool offered by Intel for accurate energy readings \cite{khan2018rapl}. For accurate measurement, we made sure that no other process is running on the devices except our evaluation workloads. We did not measure the energy consumed by private or public cloud. This is because (i) it is quite challenging to accurately measure the energy consumed by an individual task running on a cloud that offers multi-tenant services \cite{mastelic2014cloud}, and (ii) unlike cloud resources, the energy consumption of edge devices is crucial as these devices have limited battery life that depletes quicker with high energy consumption. We measured execution time using the \textit{time} keyword in our Linux bash scripts, which measures the time consumed in the execution of each phase such as data generation and data processing. Finally, we used Dstat\footnote{https://linux.die.net/man/1/dstat} for measuring the consumption of resources - CPU, RAM, disk read, and disk write.

\begin{table}[t]
	%\begin{threeparttable}
	\caption{Experimental scenarios.}\label{tab:scenarios}
	\centering
	\scriptsize
	%\vspace{-3mm}
	\begin{tabular}{p{1.2cm} p{1.7cm} p{1.8cm} p{1.7cm}}
		\hline
		\textbf{Scenario\#}      &\textbf{Client}             &\textbf{Server} &\textbf{Concept}    \\\hline\hline   
		Scenario 1      &RPi                              &RPi                    &Non-offloading         \\   	
		Scenario 2      &RPi                              &Edge node  &Offloading       \\
		Scenario 3      &RPi               &Edge server node &Offloading \\
		Scenario 4      &RPi                         & Private cloud &Offloading\\
		Scenario 5      &RPi                         & Public cloud &Offloading\\\hline
		Scenario 6      &Edge node                   &Edge node &Non-offloading \\
		Scenario 7      &Edge node                   &Edge server node  &Offloading\\
		Scenario 8      &Edge node                   &Private cloud  &Offloading\\
		Scenario 9     &Edge node                   &Public cloud  &Offloading\\\hline
		Scenario 10     &Edge server node            &Edge server node &Non-offloading \\
		Scenario 11     &Edge server node            &Private cloud  &Offloading\\
		Scenario 12     &Edge server node                        &Public cloud &Offloading \\\hline
	\end{tabular}
	%//\end{threeparttable}
	\vspace{-3mm}
\end{table}

\subsection{Evaluation Scenarios}\label{scenarios}
\begin{figure}[t]
\captionsetup{justification=centering}
\centering
  \begin{subfigure}{0.48\textwidth}
  \centering
  \includegraphics[width=\linewidth]{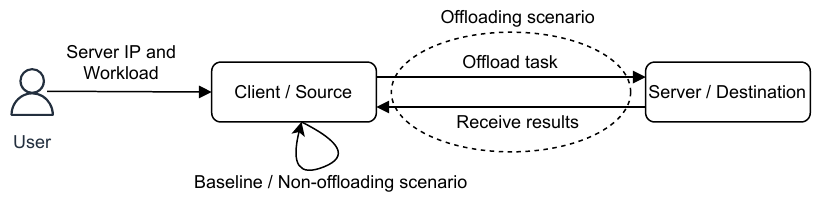}
%   \caption{Flink}
  %\vspace{5.00mm}
  %\label{bandwidth}
\end{subfigure}
\caption{Modular components for executing the scenarios.}\label{modular_components}
\label{components}
\vspace{-0.5 em}
\end{figure}

In an edge-cloud environment, tasks are either executed locally on the edge device or offloaded to a resource-rich cloud or nearby server.  Given the possibility of several scenarios, it is crucial to clearly define the evaluation scenarios with respect to the evaluation goal. In our study, we considered 12 experimental scenarios, which are illustrated in Table \ref{tab:scenarios}. These scenarios differ from each other in terms of client (node offloading the task) and server (node receiving the task). The tasks are offloaded from resource-constrained device to resource-rich device. Additionally, we also considered executing the task on the client node itself. We term such scenarios as non-offloading, which are treated as baseline for comparison with the offloading scenarios. As shown in Fig. \ref{modular_components}, for the evaluation of offloading scenarios, the client node receives the server node IP and workload as input from the user. Once input is received, all probes for the measurement of the evaluation metrics are activated. Immediately after, the client node generates the dataset for the specified workload and offloads it to the provided server node.  Evaluation metrics are continuously measured during the time the server node process the data. Once the server completes the data processing job, the results are sent back to the client and the measurement process stops. Finally, the measurements are written to a file. In case of non-offloading scenarios, the client node itself executes the workload and measures the evaluation metrics.

\subsection{Hardware Infrastructure}
The energy evaluation results directly depend upon the hardware used during the evaluation. For instance, the energy consumed during the execution of a task by RPI model 4 maybe different from that of RPI model 3. Hence, the hardware infrastructure should be clearly specified to help others understand and replicate the evaluation study. The hardware infrastructure used in our study consists of five types of nodes as described in Table \ref{infrastructure}.  \textit{RPI:} We have used a single RPI (Model 4) equipped with 4 core CPU, 8 GB RAM, and 128 GB storage. \textit{Edge Node:} The laptop, referred as edge node, is equipped with 12 core CPU, 16 GB RAM, and 256 GB storage. \textit{Edge Server Node:} The high-end server, referred as edge server node, has 8 core CPU, 32 GB RAM, and 1000 GB storage. \textit{Private Cloud:} We have used OpenStack\footnote{https://www.openstack.org/} as a private cloud center located at the University of Adelaide. The private cloud consisted of 16 nodes with each node having 1 core vCPU, 2 GB RAM, and 10 GB disk storage. \textit{Public Cloud:} We have used Google Cloud\footnote{https://cloud.google.com/} as our public cloud center located in Sydney region. Similar to private cloud, the public cloud consists of 16 nodes with each node having 1 core vCPU, 2 GB RAM, and 10 GB disk. 

\begin{table}[t]
	\caption{A summary of our infrastructure setup.}\label{infrastructure}
	\centering
	\scriptsize
	%\vspace{-3mm}
	\begin{tabular}{p{2.3cm} p{0.8cm} p{1cm} p{1cm} p{1cm}}
		\hline
		\textbf{Computing node}     &\textbf{Number}     &\textbf{CPU (cores)}         &\textbf{RAM}     &\textbf{Disk}     \\\hline\hline
		RPi                &1        &4        &8 GB   &128 GB          \\
		Edge Node          &1         &12        &16 GB   &256 GB    \\ 
		Edge Server Node   &1         &8        &32 GB  &1000 GB                \\
		Private Cloud VM         &16       &1        &2 GB  &10 GB                \\ 
		Public Cloud VM          &16       &1         &2 GB  &10 GB                \\ \hline
	\end{tabular}
	\vspace{-2 em}
\end{table}

\begin{figure*}
\captionsetup{justification=centering}
\centering
\begin{subfigure}{.24\textwidth}
  \centering
  \includegraphics[width=\linewidth]{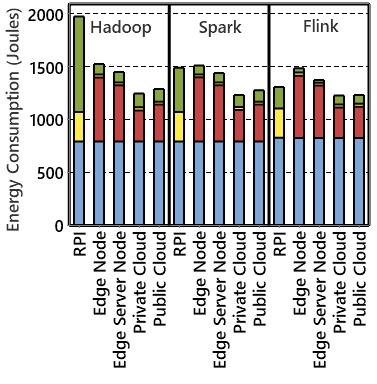}
  \caption{Grep}
  %\vspace{5.00mm}
  \label{cloud-burst-a}
\end{subfigure}%
\begin{subfigure}{.24\textwidth}
  \centering
  \includegraphics[width=\linewidth]{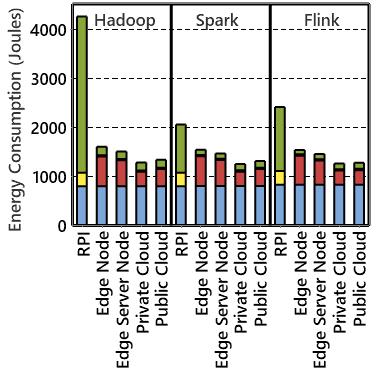}
  \caption{Word Count}
  %\vspace{5.00mm}
  \label{cloud-burst-b}
\end{subfigure}%
\begin{subfigure}{.24\textwidth}
  \centering
  \includegraphics[width=\linewidth]{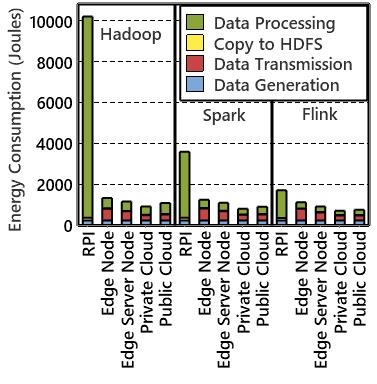}
  \caption{K-Means}
  %\vspace{5.00mm}
  \label{cloud-burst-c}
\end{subfigure}
\begin{subfigure}{.24\textwidth}
  \centering
  \includegraphics[width=\linewidth]{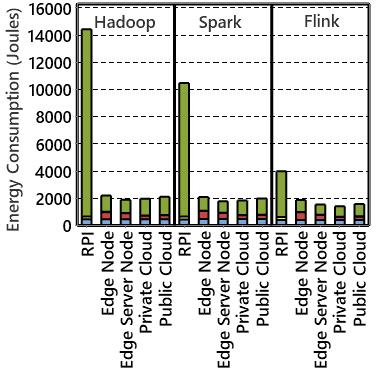}
  \caption{Page Rank}
  %\vspace{5.00mm}
  \label{cloud-burst-c}
\end{subfigure}
\caption{Energy consumption during task offloading from \textbf{RPI} to various destinations as specified on x-axis.}\label{energy_rpi}
\label{Cloud-bursting}
\vspace{-2 em}
\end{figure*}

\begin{figure}
\captionsetup{justification=centering}
\centering
\begin{subfigure}{.16\textwidth}
  %\centering
  \includegraphics[width=\linewidth]{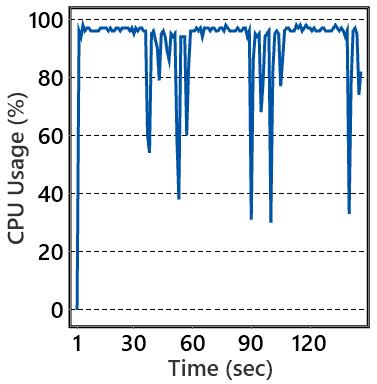}
  \caption{Data Generation}
  %\vspace{5.00mm}
  \label{cloud-burst-a}
\end{subfigure}\hfill%
\begin{subfigure}{.16\textwidth}
  %\centering
  \includegraphics[width=\linewidth]{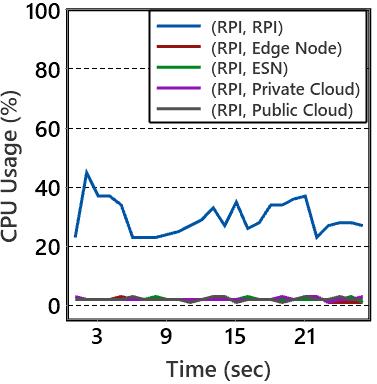}
  \caption{Data Copying}
  %\vspace{5.00mm}
  \label{cloud-burst-b}
\end{subfigure}%
\begin{subfigure}{.16\textwidth}
  %\centering
  \includegraphics[width=\linewidth]{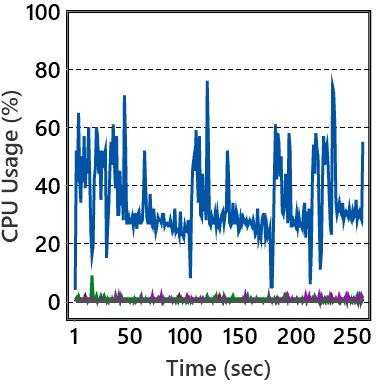}
  \caption{Data Processing}
  %\vspace{5.00mm}
  \label{cloud-burst-b}
\end{subfigure}%
\caption{CPU Usage of \textbf{RPI} during data generation, data copying, and data processing for Scenarios 1 - 5 (Table \ref{tab:scenarios}).}
\label{rpi-cpu}
\vspace{-1.5 em}
\end{figure}

\begin{figure}
\captionsetup{justification=centering}
\centering
\begin{subfigure}{.24\textwidth}
  %\centering
  \includegraphics[width=\linewidth]{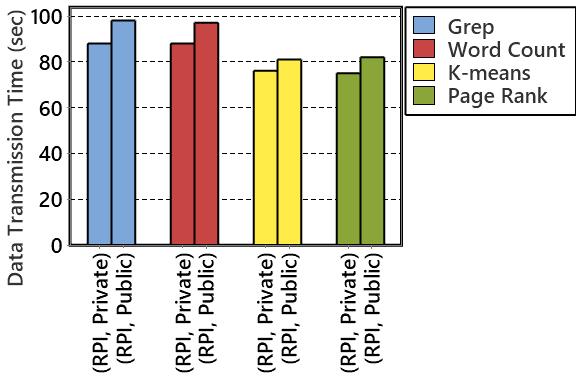}
  \caption{Data Transmission Time}
  %\vspace{5.00mm}
  \label{cloud-burst-a}
\end{subfigure}\hfill%
\begin{subfigure}{.24\textwidth}
  %\centering
  \includegraphics[width=\linewidth]{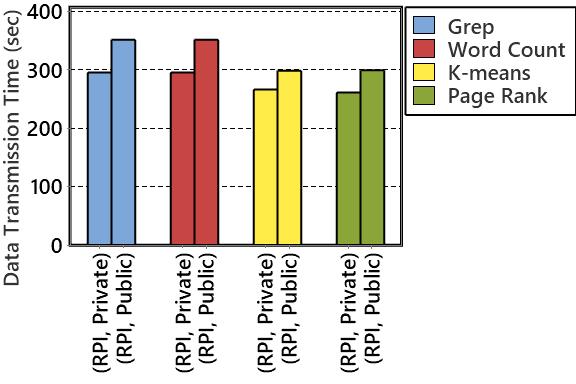}
  \caption{Data Transmission Energy}
  %\vspace{5.00mm}
  \label{cloud-burst-b}
\end{subfigure}%
\caption{Time and energy consumed in transferring data from \textbf{RPI} to private and public cloud.}\label{private_public_time_energy}
\label{Cloud-bursting}
\vspace{-5mm}
\end{figure}

\section{Results}\label{Results}
In this section, we present and compare the energy consumption of Hadoop, Spark, and FLink deployed on RPI, edge node, and edge server node.

%\textcolor{blue}{Hadoop energy usage is better for batch workload due to better data sorting approach and lower CPU usage. Hadoop is not as energy efficient as Spark and Flink for iterative workloads because iterative workloads are I/O bound. Hadoop use less CPU especially during reduce phase thats why its energy usage is low for batch workloads. Disk read and write is a bottleneck for Hadoop and Spark as Flink uses different data flow model because Flink architecture is for streaming while Hadoop and Spark are based on batch processing engines. BDEv 3.0 paper}
%\textcolor{red}{how the use of WAN impacts the performance - this should be explained with solid results (R2 - CCGRID} \textcolor{blue}{Spark takes longer time in initialization because when the execution of a Spark job starts, it creates Spark context within which Spark runs, however, creating Spark context is a very slow process. This time can be decreased through paramter tuning or reusing Spark context, which is outside the scope of this work (2) Spark uses RDD to cache intermediate data across a set of nodes between iterations to efficiently support iterative workloads (Spark vs Flink paper)}

\subsection{Energy Consumption of RPI}
Fig. \ref{energy_rpi} presents the energy consumed by the RPI during the execution of the workload either on-premise using RPI or offloading the execution to one of the four destinations - edge node, edge server node, private cloud, or public cloud. For all platforms and workloads, we observe that as compared to the offloading scenarios, the energy consumption is significantly higher in the non-offloading scenario where the workload is executed using the RPI. This is because in the non-offloading scenario, the data is processed by the RPI itself where the RPI uses its own resources such as CPU and RAM. Fig. \ref{rpi-cpu} shows the CPU usage of RPI executing k-means using Hadoop for non-offloading and offloading scenarios. Since the CPU usage in offloading scenarios is quite low (1 - 4\%), the difference in CPU usage among various offloading scenarios is not clearly visible in Fig. \ref{rpi-cpu}. We observe that in non-offloading scenario the mean CPU utilization of the RPI is around 34.2\%. On the other hand, in the offloading scenarios, the data is processed by the node/cluster at the destination during which time the RPI is idle and waiting. That is why the CPU usage of RPI in offloading scenarios is merely 1-4\%. This implies that in order to save energy, it is feasible to offload tasks to one of the destinations. 
\begin{figure}
\captionsetup{justification=centering}
\centering
\begin{subfigure}{.20\textwidth}
  %\centering
  \includegraphics[width=\linewidth]{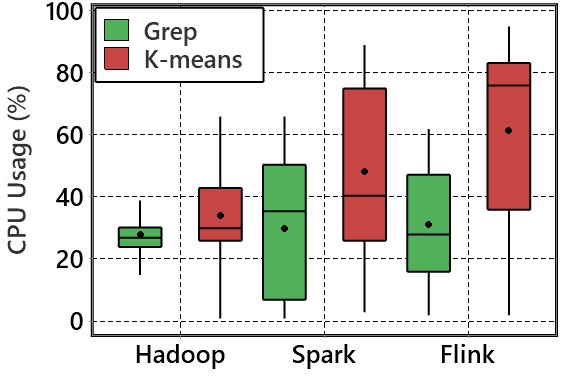}
  \caption{CPU Usage}
  %\vspace{5.00mm}
  \label{cloud-burst-a}
\end{subfigure}
\begin{subfigure}{.20\textwidth}
  %\centering
  \includegraphics[width=\linewidth]{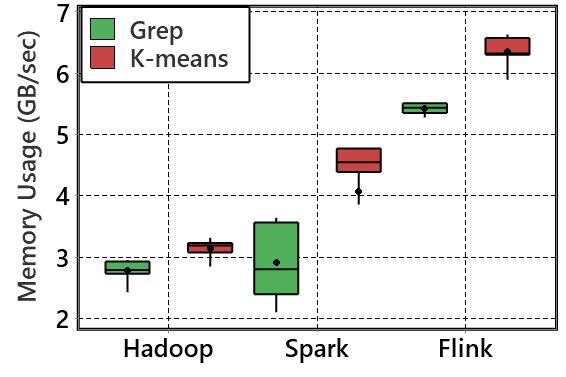}
  \caption{Memory Usage}
  %\vspace{5.00mm}
  \label{cloud-burst-b}
\end{subfigure}
\caption{CPU Usage of batch (grep) and iterative (k-means) workload for \textbf{RPI} baseline/non-offloading scenario.}
\label{cpu_batch_iterative}
\vspace{-1 em}
\end{figure}

\begin{figure}
\captionsetup{justification=centering}
\centering
\begin{subfigure}{.16\textwidth}
  %\centering
  \includegraphics[width=\linewidth]{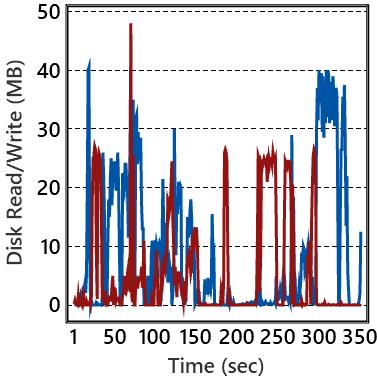}
  \caption{Hadoop}
  %\vspace{5.00mm}
  \label{cloud-burst-a}
\end{subfigure}\hfill%
\begin{subfigure}{.16\textwidth}
  %\centering
  \includegraphics[width=\linewidth]{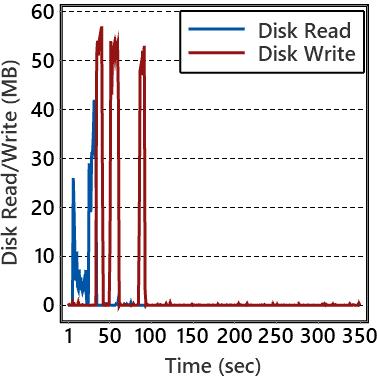}
  \caption{Spark}
  %\vspace{5.00mm}
  \label{cloud-burst-b}
\end{subfigure}%
\begin{subfigure}{.16\textwidth}
  %\centering
  \includegraphics[width=\linewidth]{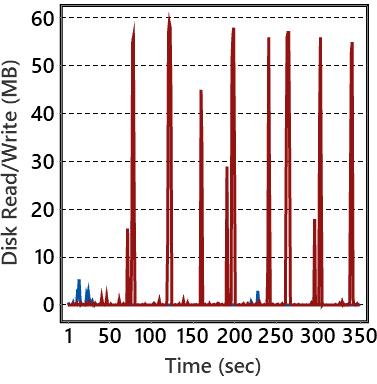}
  \caption{Flink}
  %\vspace{5.00mm}
  \label{cloud-burst-b}
\end{subfigure}%
\caption{Disk read and write for Hadoop, Spark, and Flink executing K-means on \textbf{RPI} during non-offloading scenario.}
\label{disk-usage}
\vspace{-2 em}
\end{figure} 

\begin{figure*}
\captionsetup{justification=centering}
\centering
\begin{subfigure}{.24\textwidth}
  \centering
  \includegraphics[width=\linewidth]{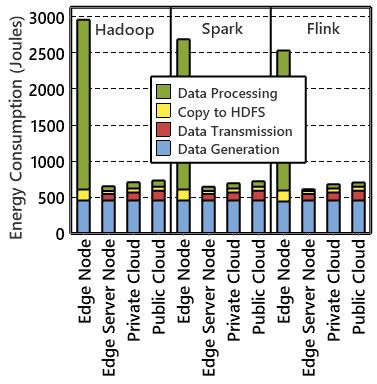}
  \caption{Grep}
  %\vspace{5.00mm}
  \label{GP-edgenode}
\end{subfigure}%
\begin{subfigure}{.24\textwidth}
  \centering
  \includegraphics[width=\linewidth]{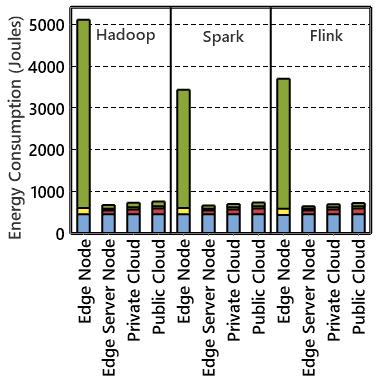}
  \caption{Word Count}
  %\vspace{5.00mm}
  \label{WC-edgenode}
\end{subfigure}%
\begin{subfigure}{.24\textwidth}
  \centering
  \includegraphics[width=\linewidth]{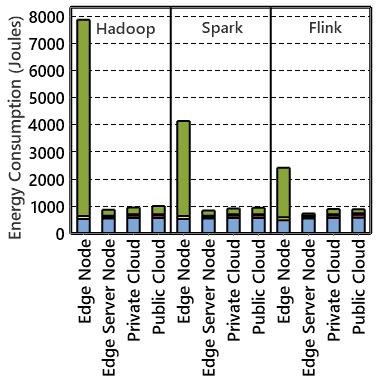}
  \caption{K-Means}
  %\vspace{5.00mm}
  \label{KM-edgenode}
\end{subfigure}
\begin{subfigure}{.24\textwidth}
  \centering
  \includegraphics[width=\linewidth]{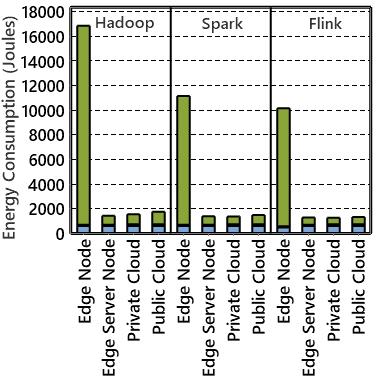}
  \caption{Page Rank}
  %\vspace{5.00mm}
  \label{PR-edgenode}
\end{subfigure}
\caption{Energy consumption during task offloading from \textbf{edge node} to various destinations as specified on x-axis}\label{edgenode_energy}
\label{energy-edge-node}
\vspace{-1.0 em}
\end{figure*}

As compared to non-offloading scenario, offloading task from RPI reduces the mean energy consumption of RPI by 51.6\%. Offloading the task to edge node, edge server node, private cloud, and public cloud reduces mean energy by 44.7\%, 49.6\%, 56.6\%, and 54.5\%, respectively. The reduction in case of offloading to edge server node as compared to edge node is slighter higher due to the difference in the energy consumed during data transmission and data processing phase. Since the bandwidth (4.1MB/s) between RPI and edge server node is higher than the bandwidth (2.7MB/s) between RPI and edge node as shown in Fig. \ref{bandwidth}, it takes lower time and energy to transmit data to edge server node. Edge server node has higher RAM and CPU power as compared to edge node as reported in Table \ref{infrastructure}, therefore, edge server node process the data quickly. As a result, RPI receives the result quicker which directly translates into lower energy consumption. Although both private and public clouds are equipped with similar resources such as RAM and CPU, the energy consumption during task offloading to private cloud is 2.2\% lower than the public cloud. This difference mainly stems from the difference in the energy consumption during data transmission phase. In our experimental setup, the distance between RPI and private cloud (in Adelaide) is around 107 meters while the distance between RPI and public cloud (in Sydney) is around 1374 KM. Hence, it takes longer time to transmit data to public cloud, which directly results in higher consumption of energy as illustrated in Fig. \ref{private_public_time_energy}. Offloading to public or private cloud as compared to edge server nodes saves more energy. This is a direct determinant of the higher bandwidth (5.8 - 6.2MB/s) between RPI and cloud as compared to 4.1MB/s bandwidth between RPI and edge server node.  
%(\textcolor{red}{figure showing bars of time and energy consumed in transferring data to private and public cloud}.\textcolor{blue}{Can we draw some boxplots for some results?}

With respect to workloads, it is evident from Fig. \ref{energy_rpi} that iterative workloads (k-means and page rank) consume quite high energy as compared to batch workloads - grep and word count. This is because the execution of batch workload requires processing the data only once while iterative workloads require multiple iterations of data processing in order to produce the final output. The energy consumed in data generation, data transmission, and copy to HDFS is almost similar across all workloads. The main difference occurs during the data processing phase where the batch workloads consumes around 79.6\% less energy as compared to the iterative workloads. Furthermore, as presented in Fig. \ref{cpu_batch_iterative}, we also observe a lower CPU usage during the execution of batch workloads as compared to iterative workloads i.e., 28.1\% vs 34.2\% for Hadoop, 29.8\% vs 48.2\% for Spark, and 31.3\% vs 61.5\% for Flink. Higher usage of CPU also contributes to the higher consumption of energy as observed for iterative workloads.

Fig. \ref{energy_rpi} also depicts the energy consumption with respect to the platforms. For baseline scenario, with the exception of one case, the energy consumption with Flink is lowest followed by Spark and Hadoop. This trend is aligned with the disk usage by these platforms. As shown in Fig. \ref{disk-usage}, Hadoop makes the most use of disk, which slows it down in terms of data processing while Flink rarely spells data to the disk, which enables it to process the data quicker. With respect to Fig. \ref{disk-usage}, the mean data written to the disk per second by Hadoop, Spark, and Flink is 10.9 MB, 3.9 MB, and 3.7 MB, respectively. Such behaviour in terms of disk usage directly impacts the execution time, which results in higher energy usage for Hadoop. Flink outclasses Spark in terms of energy consumption due to its in-built optimizer too, which enables Flink to use CPU and RAM more efficiently. For offloading scenarios, the energy consumption across the three platforms is similar. This is because irrespective of the platform deployed on the RPI, the RPI has to wait for the server to process the data and receive the results.  %(\textcolor{red}{Why the energy consumption is similar? Shouldn't Flink consumes less energy given that is faster as compared to Hadoop and Spark}

\begin{figure}
\captionsetup{justification=centering}
%\centering
\begin{subfigure}{.24\textwidth}
  %\centering
  \includegraphics[width=\linewidth]{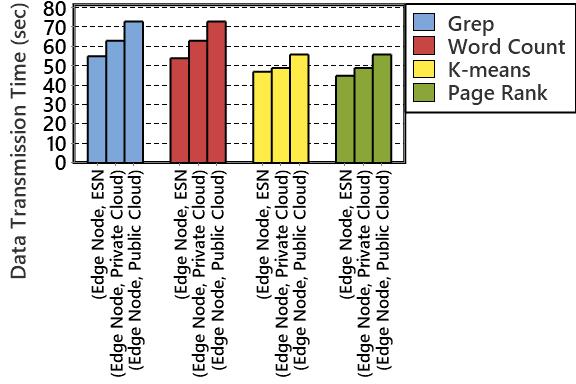}
  \caption{Data Transmission Time}
  %\vspace{5.00mm}
  \label{cloud-burst-a}
\end{subfigure}\hfill%
\begin{subfigure}{.24\textwidth}
  %\centering
  \includegraphics[width=\linewidth]{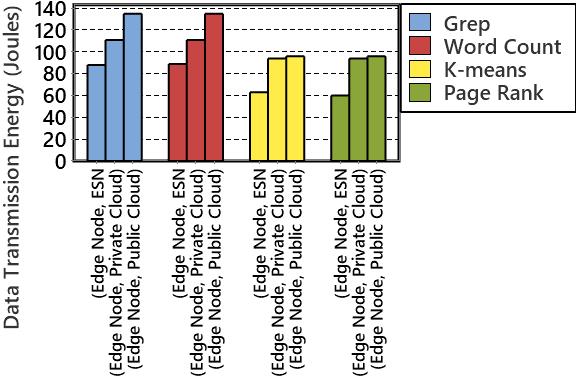}
  \caption{Data Transmission Energy}
  %\vspace{5.00mm}
  \label{cloud-burst-b}
\end{subfigure}%
\caption{Time and energy for transferring data from \textbf{edge node} to edge server node, private cloud, and public cloud}\label{energy_time_edgenode}
\label{Cloud-bursting}
%\vspace{-0.5 em}
\end{figure}

\begin{table}[t]
%\resizebox{\textwidth}{!}{%
\caption{Data processing time (in seconds) for Hadoop, Spark, and Flink deployed on \textbf{edge node}}
\centering
\begin{tabular}{l||cccc}
\hline
% Platform & \multicolumn{4}{c|}{Workload} \\ \hline
\textbf{Workload} & \multicolumn{1}{c|}{\textbf{Grep}} & \multicolumn{1}{c|}{\textbf{Word Count}} & \multicolumn{1}{c|}{\textbf{K-means}} & \textbf{Page Rank} \\ \hline \hline
Hadoop & \multicolumn{1}{c|}{47} & \multicolumn{1}{c|}{109} & \multicolumn{1}{c|}{260} & 600 \\ 
Spark & \multicolumn{1}{c|}{40} & \multicolumn{1}{c|}{75} & \multicolumn{1}{c|}{201} & 502 \\ 
Flink & \multicolumn{1}{c|}{18} & \multicolumn{1}{c|}{72} & \multicolumn{1}{c|}{62} & 421 \\ \hline
\end{tabular}
\label{time-edgenode}
\vspace{-2 em}
\end{table}

\begin{figure*}
\captionsetup{justification=centering}
\centering
\begin{subfigure}{.24\textwidth}
  \centering
  \includegraphics[width=\linewidth]{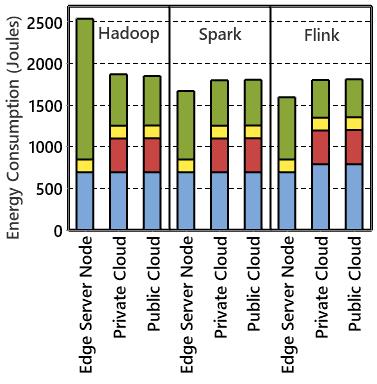}
  \caption{Grep}
  %\vspace{5.00mm}
  \label{edgserver-a}
\end{subfigure}%
\begin{subfigure}{.24\textwidth}
  \centering
  \includegraphics[width=\linewidth]{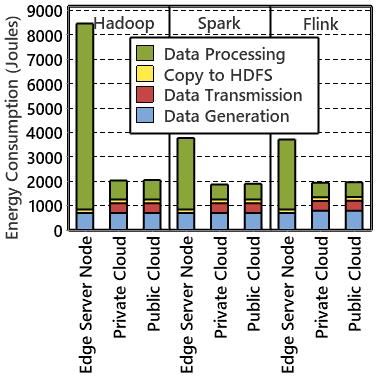}
  \caption{Word Count}
  %\vspace{5.00mm}
  \label{edgserver-b}
\end{subfigure}%
\begin{subfigure}{.24\textwidth}
  \centering
  \includegraphics[width=\linewidth]{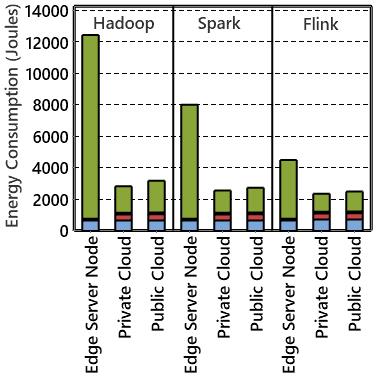}
  \caption{K-Means}
  %\vspace{5.00mm}
  \label{edgserver-c}
\end{subfigure}
\begin{subfigure}{.24\textwidth}
  \centering
  \includegraphics[width=\linewidth]{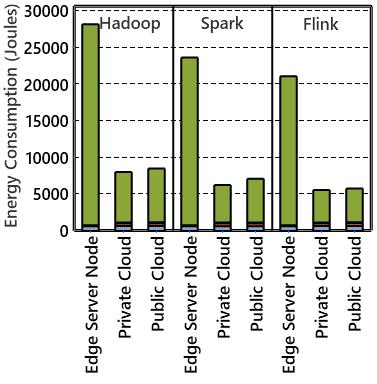}
  \caption{Page Rank}
  %\vspace{5.00mm}
  \label{edgserver-d}
\end{subfigure}
\caption{Energy consumption during task offloading from \textbf{edge server node} to various destinations specified on x-axis}
\label{energy-edgeserver}
\vspace{-1.5 em}
\end{figure*}

\subsection{\textit{Energy Consumption of Edge Node}}\label{rq1}
Fig. \ref{edgenode_energy} presents the energy consumption for the edge node. It is evident that in comparison to offloading scenarios, the energy consumption is quite high in the non-offloading scenarios where the task is executed on the edge node. This is because the edge node remains idle during data processing stage in offloading scenarios as the task is executed on the destination node such as edge server node or cloud. In the non-offloading scenario, similar to RPI, the edge node uses its own CPU and RAM to process the data. During data processing stage, we observe that the mean CPU utilization of the edge node reduces from 59.4\% in non-offloading scenario to 2.6\% in offloading scenarios (figure not included due to space limitation).

In non-offloading scenario, on average, 85.3\% energy is consumed in data processing. In offloading scenarios, the energy distribution significantly differs between batch and iterative workloads. For batch workloads (grep and word count), on average, 65.7\% energy is consumed in data generation, 16.2\% in data transmission, 7.5\% in copying data to HDFS, and 10.8\% in data processing. For iterative workloads (k-means and page rank) that are compute-heavy, the distribution of mean energy consumption is as follows - 51.8\% data generation, 7.8\% data transmission, 3.8\% copying data to HDFS, and 36.4\% data processing. 
In terms of saving energy, the best option for edge node is to offload the task to the edge server node as least amount of energy is consumed when tasks are offloaded to the edge server node. Whilst the data generation is same for offloading to edge server node and private/public cloud, the difference is created by the energy consumed during data transmission. We observe that the energy consumed in data transmission is directly proportional to the distance between client and server. For instance, the edge node is located at a distance of around 107 meters from our private cloud and around 1375 KM from the public cloud. Thereby, the data transmission energy is lower for transferring data to the private cloud as compared to the public cloud as illustrated in Fig. \ref{energy_time_edgenode}. Fig. \ref{energy_time_edgenode} shows that energy consumed by edge node to transmit data to edge server node is 35.1\% and 27.4\% less than the energy consumed for transmitting data to the public and private cloud, respectively. Similarly, the mean energy consumed by edge node to transfer data to private cloud is 9.9\% less than the energy consumed during data transmission to public cloud. 

%\textcolor{blue}{(1) Understanding the exact factors impacting performance is challenging because (1) tasks use multiple resource simultaneously (ii) even a single jobs consists of multiple tasks that run in parallel - USENIX 2015 Kay Ousterhout (2) In Hadoop, a lot of time is spent on data shuffling, during which the workers communicate with the master to find out where the data is located (2) Jobs botleneck on the disk before bottlenecking on the network becuase the data transferred via network is only a subset of the data transfered to the disk (USENIX 2015 Kay ousterhout) (3) the number of tasks run in parallel for each framework is different by default. for example, 23 in Hadoop, 2 in Spark, and 16 in Flink in our setup. So it also impacts the performance (USENIX 2015 Kay ousterhout) (4) if the CPU cores are much higher as compared to the number/capacity of disk, a lot of time will be spent where CPU waits for the disk to read and write the data. 1:3 of disk to cpu cores is ideal}

The energy consumption across the three platforms is similar in offloading scenarios. This is as expected because in offloading scenarios, the edge node remains idle during data processing stage irrespective of which platform is deployed. Moreover, the data transmission energy dominates the energy consumed during that idle stage. In non-offloading scenario, Flink consumes least energy followed by Spark while Hadoop consumed the most energy. This is largely a determinant of the data processing time as shown in Table \ref{time-edgenode} for the three platforms. Since Flink takes minimum time to process the data, it consumes less energy. On the other hand, Hadoop being a disk-based platform, takes comparatively more time, which leads to higher energy consumption. The execution time largely relates to disk I/O, which is quite high in case of Hadoop as compared to Spark and Flink. According to our results for edge node, Hadoop reads and writes around 3.5 MB/sec and 3.7 MB/sec data to the disk. The disk read and write rate for Spark is 1.8 MB/sec and 1.7 MB/sec and for Flink is 1.2 MB/sec and 1.1 MB/sec.

%\begin{figure}
%\captionsetup{justification=centering}
%\centering
%\begin{subfigure}{.30\textwidth}
  %\centering
%  \includegraphics[width=\linewidth]{figs/Energy/time-laptop-baseline.jpg}
  %\caption{Data Transmission Time}
  %\vspace{5.00mm}
%  \label{cloud-burst-a}
%\end{subfigure}
%\vspace{-1.5 em}
%\caption{Data processing time for Hadoop, Spark, and Flink deployed on \textbf{edge node}}
%\label{time-laptop-baseline}
%\label{Cloud-bursting}
%\end{figure}

The energy consumption of the edge node in case of offloading tasks to private and public cloud is similar. This is because (i) the data generation phase accounting for 50-65\% energy is similar for both private and public clouds (ii) the data processing phase accounting for 10-35\% energy is almost same in both clouds as both the clouds have the same resource configuration. Therefore, the data processing takes approximately similar time in private and public clouds as illustrated in Fig. \ref{energy-edge-node}. The minor difference is caused by the data transmission phase where transmitting data to public cloud, being farther, consumes 9.9\% more energy as compared to private cloud. Comparing Fig. \ref{GP-edgenode} and Fig. \ref{WC-edgenode} for batch workloads with Fig. \ref{KM-edgenode} and Fig. \ref{PR-edgenode} for iterative workloads shows that iterative workloads consumes far more energy than batch workloads on edge node as was the case for RPI too. 

\subsection{Energy Consumption of Edge Server Node}
Fig. \ref{energy-edgeserver} shows the energy consumed by edge server node as it offloads tasks to various computing options or execute the task itself. Similar to RPI and egde node, the energy consumed in the non-offloading scenarios is quite high as compared to offloading scenarios for word count, k-means, and page rank. However, this is not true for grep with Spark and Flink due to two reasons. First, grep is a light-weight workload so the energy consumed during the data processing stage using edge server node and using resourceful private or public cloud is not too much different. Furthermore, the data transmission energy contributes significantly to the energy consumed in offloading workload to the public or private cloud. In comparision to Spark and Flink, executing grep using Hadoop on edge server node consumes more energy as compared to offloading scenarios. This is because Hadoop consumes far more energy during data processing, which cannot be dominated by the data transmission energy. 

\begin{figure}
\captionsetup{justification=centering}
%\centering
\begin{subfigure}{.16\textwidth}
  %\centering
  \includegraphics[width=\linewidth]{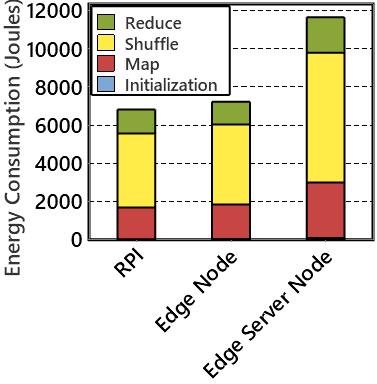}
  \caption{Hadoop}
  %\vspace{5.00mm}
  \label{cloud-burst-a}
\end{subfigure}\hfill%
\begin{subfigure}{.16\textwidth}
  %\centering
  \includegraphics[width=\linewidth]{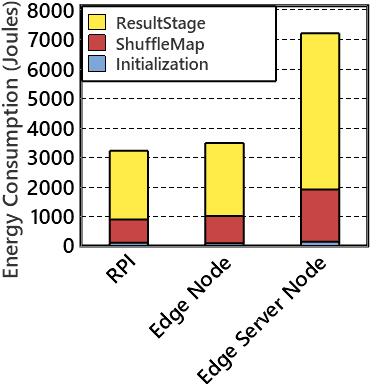}
  \caption{Spark}
  %\vspace{5.00mm}
  \label{cloud-burst-b}
\end{subfigure}%
\begin{subfigure}{.16\textwidth}
  %\centering
  \includegraphics[width=\linewidth]{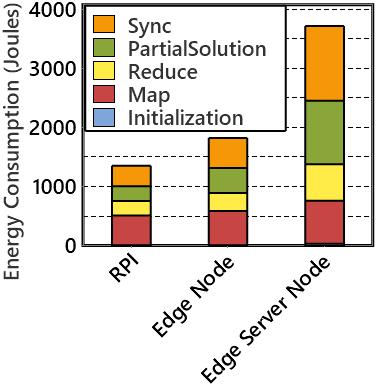}
  \caption{Flink}
  %\vspace{5.00mm}
  \label{cloud-burst-b}
\end{subfigure}%
\caption{Energy consumed in each stage during execution of K-means on \textbf{RPI}, \textbf{edge node}, \textbf{edge server node} in non-offloading/baseline scenario}
\label{stages}
\vspace{1 em}
\end{figure}

Fig. \ref{stages} shows the energy consumed in the execution of various phases of data processing for baseline scenario. On average, the initialization of the platform consumes only 0.68\%, 2.72\%, and 0.94\% of the total energy for Hadoop, Spark, and Flink, respectively. Hence, it can be asserted that the initialization phase does not dominate the actual processing stages in our study. Spark consumes higher energy in initialization because when Spark job starts, it creates SparkContext, which is a slow process. Hadoop consumes comparatively higher energy in shuffle stage. This is attributed to the synchronization barrier (each thread waiting for the antecedent thread to finish writing) that Hadoop faces while writing intermediate data to the disk during shuffle stage.

%Edge server node consumes 1587 - 2530 Joules to execute Grep, which is the most light-weight workload. The distribution of the energy consumption of edge server node for grep is quite different from the edge node. For instance, in baseline scenario, edge server node consumes 40 - 60\% energy in data processing unlike 70 - 80\% consumed by edge node. This happens due to \textcolor{red}{mention reason here}. In case of Spark and Flink with grep, it is more energy-efficient to execute the task locally in case of offloading to cloud. Although the energy consumed in data processing locally is higher than the private and public cloud, the data transmission energy makes the difference, which is 0 in case of local processing as compared to 400 - 500 Joules consumed in case of offloading to the cloud. The same is not observed for Hadoop because the data processing energy locally is too high (1400 Joules) as compared to 700 - 800 Joules consumed in case of offloading to the cloud. 

\begin{figure}
\captionsetup{justification=centering}
%\centering
\begin{subfigure}{.24\textwidth}
  %\centering
  \includegraphics[width=\linewidth]{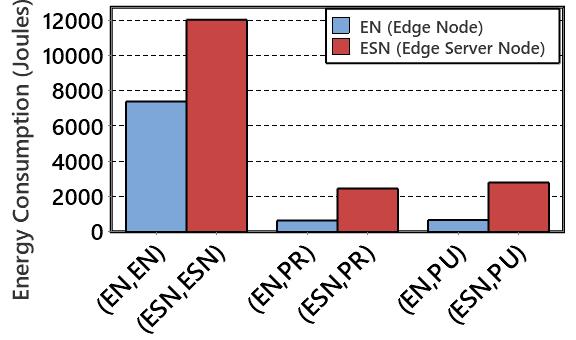}
  \caption{CPU Energy}
  %\vspace{5.00mm}
  \label{cloud-burst-a}
\end{subfigure}%
\begin{subfigure}{.24\textwidth}
  %\centering
  \includegraphics[width=\linewidth]{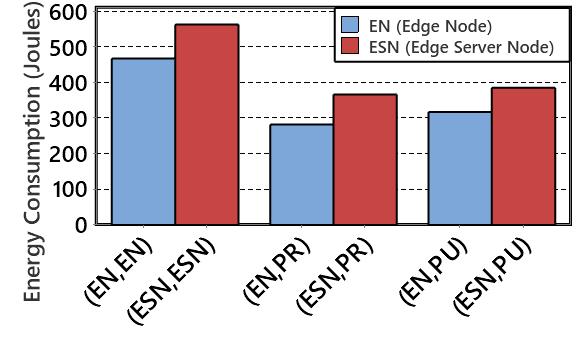}
  \caption{RAM Energy}
  %\vspace{5.00mm}
  \label{cloud-burst-b}
\end{subfigure}%
\caption{Energy consumed by CPU and RAM of \textbf{Edge Node} (EN) and \textbf{Edge Server Node} (ESN) with Hadoop executing K-means in various scenarios. PR and PU denotes private and public cloud, respectively}
\label{cpu-ram-energy}
\vspace{-1.5 em}
\end{figure}

\begin{figure}[t]
\captionsetup{justification=centering}
\centering
\begin{subfigure}{.17\textwidth}
  \centering
  \includegraphics[width=\linewidth]{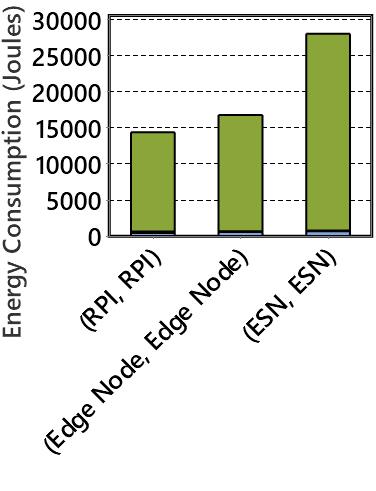}
  \caption{Non-offloading}
  %\vspace{5.00mm}
  \label{rpi-edgenode-edgeserver-energy-a}
\end{subfigure}%
\begin{subfigure}{.32\textwidth}
  \centering
  \includegraphics[width=\linewidth]{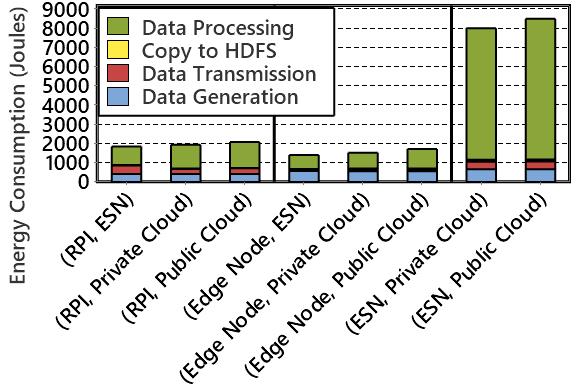}
  \caption{Offloading}
  %\vspace{5.00mm}
  \label{rpi-edgenode-edgeserver-energy-b}
\end{subfigure}
\caption{Energy consumed by \textbf{RPI}, \textbf{Edge Node}, and \textbf{Edge Server Node} (ESN) during non-offloading (local execution) and offloading scenarios for Hadoop executing Page Rank.}
\label{rpi-edgenode-edgeserver-energy}  
\vspace{-1 em}
\end{figure}

\begin{figure}
\captionsetup{justification=centering}
\centering
\begin{subfigure}{.17\textwidth}
  \centering
  \includegraphics[width=\linewidth]{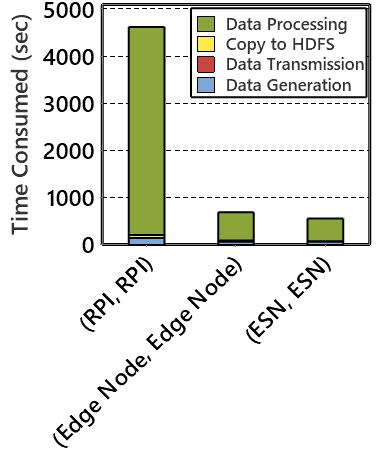}
  \caption{Non-offloading}
  %\vspace{5.00mm}
  \label{rpi-edgenode-edgeserver-time-a}
\end{subfigure}%
\begin{subfigure}{.32\textwidth}
  \centering
  \centering
  \includegraphics[width=\linewidth]{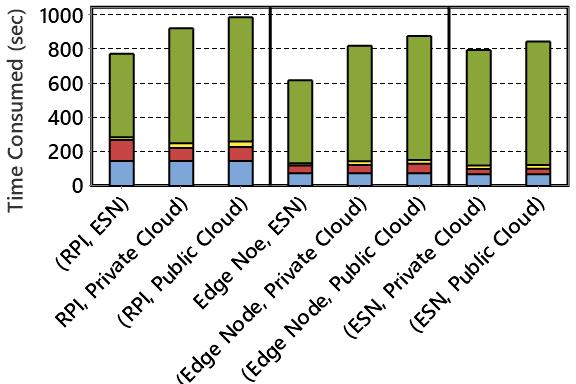}
  \caption{Offloading}
  %\vspace{5.00mm}
  \label{rpi-edgenode-edgeserver-time-b}
\end{subfigure}
\caption{Time consumed by \textbf{RPI}, \textbf{Edge Node}, and \textbf{Edge Server Node} (ESN) during non-offloading (local execution) and offloading scenarios for Hadoop executing Page Rank.}
\label{rpi-edgenode-edgeserver-time}  
\vspace{-1 em}
\end{figure}

The mean energy consumed by the edge server node in task offloading to the public cloud is 5.56\% higher than the energy consumed in offloading to private cloud. The key difference is in energy consumed during data transmission. Since private cloud is located closer (107 meters) to the edge server node as as compared to public cloud (1375 KM), it takes more energy to transfer data to the public cloud. Except grep, offloading tasks from edge server node to the cloud on average saves 74.1\% energy for Hadoop, 62.9\% for Spark, and 55.5\% for Flink. Hence, it can be concluded that offloading computationally-heavy tasks to the cloud from edge server nodes leads to saving significant amount of energy. Copying data to HDFS at the destination nodes consumes around 107-153 Joules, however, this proportion of energy is not visible in Fig. \ref{edgserver-c} and Fig. \ref{edgserver-d} due to the large chunk of energy consumed in data processing. 

%Generally, edge server node consumes higher energy as compared to edge node. As an example, for executing word count, edge server node consumes almost double the energy consumed by edge node. Fig. 7 shows the energy consumed by edge node and edge server node in each phase. It is evident that the energy consumption for data generation is almost similar. However, edge node outclasses edge server node in terms of energy efficiency for the remaining three tasks - data transmission, copy to HDFS, and data processing. 

%\textcolor{blue}{Resource utilization of edge node and edge server node - is there any difference? If there is, how that difference relates to the energy consumption}

\subsection{Energy Efficiency - RPI vs Edge Node vs Edge Server Node}
Fig. \ref{rpi-edgenode-edgeserver-energy} compares the energy consumption of client nodes including RPI, edge node, and edge server node for non-offloading and offloading scenarios. In non-offloading scenario, RPI consumes the least amount of energy while edge server node consumes the most amount of energy. This trend is directly linked with the resources available with each device as presented in Table \ref{infrastructure}. RPI being the device with limited resources consumes less energy while edge server node having more resources (CPU and RAM) consumes most energy. Whilst RPI consumes least energy, it takes the longest time to complete the job as depicted in Fig. \ref{rpi-edgenode-edgeserver-time}. In other words, RPI is energy-efficient while edge server node is time-efficient. Hence, we observe a trade-off situation where the device consuming least energy takes longest execution time and vice versa. Moreover, we also observe that around 95.4\% energy is consumed during data processing in non-offloading scenario.

%\begin{figure*}
%\captionsetup{justification=centering}
%\centering
%\begin{subfigure}{.17\textwidth}
%  \centering
%  \includegraphics[width=\linewidth]{figs/Energy/rpi vs laptop vs energy server-baseline.jpg}
%  \caption{Non-offloading (Energy)}
  %\vspace{5.00mm}
%  \label{edgserver-a}
%\end{subfigure}%
%\begin{subfigure}{.17\textwidth}
%  \centering
%  \includegraphics[width=\linewidth]{figs/Energy/rpi-en-esn-time-baseline.jpg}
%  \caption{Non-offloading (Time)}
  %\vspace{5.00mm}
%  \label{edgserver-b}
%\end{subfigure}%
%\begin{subfigure}{.32\textwidth}
%  \centering
%  \includegraphics[width=\linewidth]{figs/Energy/rpi vs laptop vs energy server-offloading.jpg}
%  \caption{Offloading (Energy)}
  %\vspace{5.00mm}
%  \label{edgserver-c}
%\end{subfigure}
%\begin{subfigure}{.32\textwidth}
%  \centering
%  \includegraphics[width=\linewidth]{figs/Energy/rpi-en-esn-time-offloading.jpg}
%  \caption{Offloading (Time)}
  %\vspace{5.00mm}
%  \label{edgserver-d}
%\end{subfigure}
%\caption{Energy and time consumed by \textbf{RPI}, \textbf{Edge Node}, and \textbf{Edge Server Node} (ESN) during non-offloading (local execution) and task offloading scenarios for Hadoop executing Page Rank.}
%\label{rpi-edgenode-edgeserver}  
%\vspace{-2 em}
%\end{figure*}

With regards to offloading scenarios, edge node consumes 26.2\% less energy than RPI during task offloading to edge server node. This difference comes from data transmission energy, where edge node consumes only 60 joules as compared to 445 joules consumed by RPI. This difference in data transmission energy is a result of the difference in the bandwidths (4.1 MB/s between RPI and edge server node vs 8.6 MB/s between edge node and edge server node) as shown in Fig. \ref{bandwidth}. Same goes true between RPI and edge node energy consumption for task offloading to private and public cloud, where the energy consumed by RPI is higher than edge node. During task offloading to private and public cloud, the energy consumed by edge node is the lowest while the energy consumed by edge server node is the highest. The key difference stems from the energy consumed in data processing phase where edge server node consumes 5 times more energy than edge node. While both nodes wait for the destination node to execute the task during these phases, wait by edge server node is more costly in terms of energy as compared to edge node. Such is illustrated in Fig. \ref{cpu-ram-energy}, where it can be observed that both CPU and RAM of edge server node consumes higher energy as compared to the CPU and RAM of the edge node. This is because edge server node is equipped with higher CPU and RAM capacity as compared to the edge node (Table \ref{infrastructure}).

\section{Conclusion}\label{conclusion}
In this paper, we proposed a framework for evaluating the energy consumption of distributed data processing platforms deployed in an integrated edge-cloud environment. We then used the proposed framework to evaluate the energy consumption of the three most widely used platforms (i.e., Hadoop, Spark, and Flink) using 12 scenarios. Our results reveal that (i) Flink is most energy-efficient followed by Spark, and Haodop is found to be the least energy-efficient (ii) offloading data processing tasks from resource-constraint to resource-rich device reduces mean energy consumption by 55.2\% (iii) the bandwidth, distance between client and server, and their computational powers are key factors to be considered during energy assessment of the distributed data processing platforms. These findings should not be taken as the final word on the energy consumption of these platforms. This is because our study only uses a few workloads and very specific infrastructure during evaluation. As the workloads or infrastructure (e.g., connection type and hardware devices) change, we expect the energy consumption will change accordingly. Therefore, the key takeaway from this work is how to evaluate the energy consumption of the distributed data processing platforms in edge cloud environment so that researchers and practitioners can identify and exploit areas for reducing energy consumption. In future, we plan to extend and leverage the proposed framework for investigating the impact of tuning parameters (e.g., replication factor, memory allocation, and data compression) on the energy consumption of the distributed data processing platforms.
%\newpage
% \end{thebibliography}

\bibliographystyle{IEEEtran}
\bibliography{IEEEabrv,SHagun.bib}
% \begin{thebibliography}{IEEEexample.bib}

% % \bibitem{}

% \end{thebibliography}
% \vspace{12pt}
% \color{red}
% IEEE conference templates contain guidance text for composing and formatting conference papers. Please ensure that all template text is removed from your conference paper prior to submission to the conference. Failure to remove the template text from your paper may result in your paper not being published.

\end{document}